\documentclass[oneside]{article}
\usepackage[T1]{fontenc}
\usepackage{authblk}
\usepackage{float}
\usepackage{amsmath}
\usepackage{graphicx}
\usepackage{xfrac}
\usepackage[english]{babel}
\usepackage{lmodern}
\usepackage[T1]{fontenc}
\usepackage[latin9]{inputenc}
\PassOptionsToPackage{normalem}{ulem}
\usepackage{ulem}
\usepackage{mathtools}
\usepackage{graphicx}
\usepackage{esint}
\usepackage[unicode=true,
 bookmarks=false,
breaklinks=false,pdfborder={0 0 1},backref=false,colorlinks=false]
 {hyperref}
\usepackage[a4paper, total={210mm,297mm},margin=2.0cm]{geometry}
\usepackage{breakurl}
\usepackage[normalem]{ulem}
\usepackage{color}
\usepackage{bm}

\title{Microscale modelling of dielectrophoresis assembly processes}

\author[1,2]{Adriano Tiribocchi}
\author[2]{Andrea Montessori}
\author[2]{Marco Lauricella\thanks{Electronic address: \texttt{marco.lauricella@cnr.it}; Corresponding author}}
\author[1,2]{Fabio Bonaccorso}
\author[3,4,5]{Keith A. Brown}
\author[1,2]{Sauro Succi}

\affil[1]{Center for Life Nanoscience at la Sapienza, Istituto Italiano di Tecnologia, viale Regina Elena 295, 00161, Rome, Italy}
\affil[2]{Istituto per le Applicazioni del Calcolo CNR, Via dei Taurini 19, 00185 Rome, Italy}
\affil[3]{Department of Mechanical Engineering, Boston University, Boston, MA 02215, USA}
\affil[4]{Division of Materials Science \& Engineering, Boston University, Boston, MA 02215, USA}
\affil[5]{Physics Department, Boston University, Boston, MA 02215, USA}

\date{\today}

\begin{document}

\maketitle

\begin{abstract}
This work presents a microscale approach for simulating the dielectrophoresis (DEP) assembly of polarizable particles under an external electric field. The model is shown to capture interesting dynamical and topological features, such as the formation of chains of particles and their incipient aggregation into hierarchical structures. A quantitative characterization in terms of the number and size of these structures is also discussed. This computational model could represent a viable numerical tool to study the mechanical properties of particle-based hierarchical materials and suggest new strategies for enhancing their design and manufacture.
\end{abstract}

\section{Introduction}

\begin{figure}[!h]
\centering
\includegraphics[scale=.5]{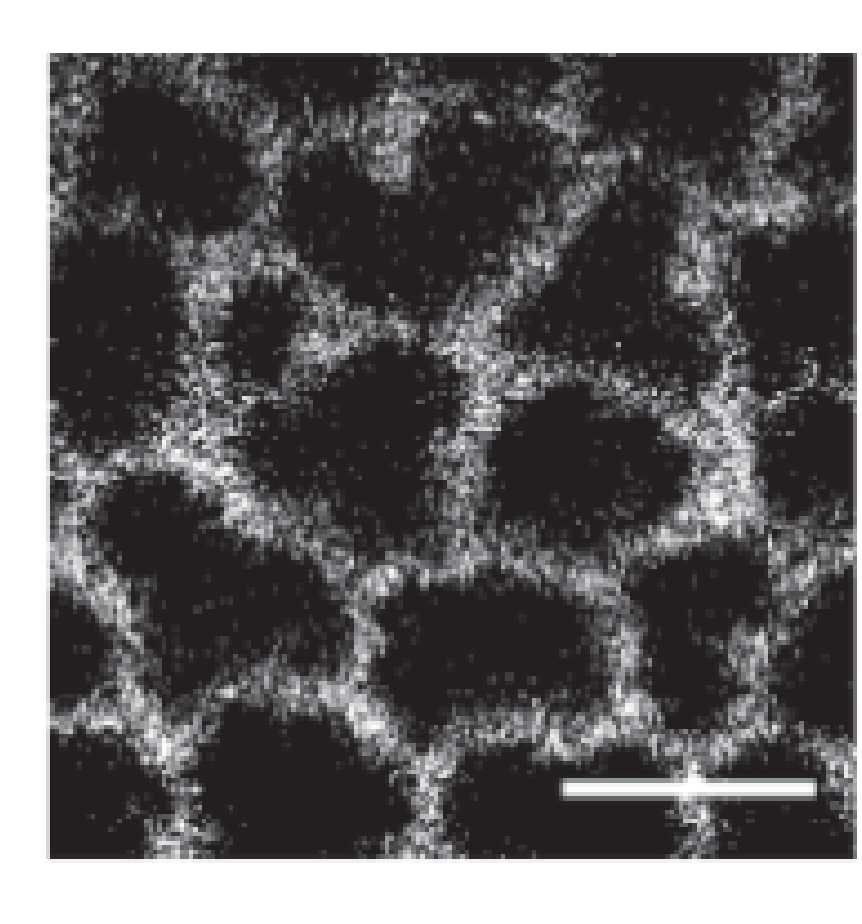}
\caption{The figure shows an experimental realization of a hierarchical cellular network made of polarizable colloidal suspension (white) surrounding voids regions (black). The figure is taken from Ref.~\cite{agarwal2009low}.}
\label{fig1}
\end{figure}

Electric fields are a handy tool for manipulating micro and nanomaterials in solutions~\cite{kozai_nat,modarres_act,lauricellaRMP}. Dielectrophoresis (DEP), i.e., the motion of polarizable objects in a non-uniform electric field, has become a vastly used technique to separate, sort, and trap systems such as cells \cite{demircan2013,salemmilani2018}, nano and microparticles \cite{keith2011}, and biomolecules \cite{modarres_act}, to name but a few. 
A remarkable feature of DEP is that it can be used to drive the assembly of significantly small materials, such as nano-colloids, thus providing, for instance, an efficient strategy to deposit dielectric and metallic nanoparticles onto electrode arrays and to control their arrangement \cite{freedman2016,wang2018}.
If coated with polymers, such systems can exhibit a hierarchical structure, like pearl chains aligned along the field lines \cite{gonzalez2015} or well-defined networks made of a percolating mesh of particle-rich walls surrounding particle-free ``voids'' (see Fig.\ref{fig1}) \cite{agarwal2009low,kumar2005new}.

The ability to assemble hierarchical structures starting from micro or nano-particles responsive to an electric field is crucial in several technological applications of modern industry, ranging from medical diagnostics and photolithography up to material science for the design of soft composites with high tunable porosity \cite{lumsdon2004two,dassanayake2000structure,furst2000dynamics,martin1998structure}. Indeed, the hierarchical DEP process may provide a highly desirable mechanism to realize large-scale ordered materials, built from a field-directed assembly mechanism dominating the Brownian motion of particles \cite{cao2019measuring}. 

Previous simulations, including dipolar and van der Waals interactions among spherical particles, have found several variants of structures (linear aggregates, droplets, columns) resulting from the phase separation into regions of a high and low density of colloidal particles~\cite{richardi2008self,stevens1995phase,tao1994simulation}. However, within these computational approaches, a model of polarizable colloidal particles has not been considered so far.

Here we present a numerical scheme aimed at describing the assembly process of nano-particles capable of forming instantaneous dipoles under the effect of the surrounding electric field. The model is based on a classical version of Drude oscillators \cite{lamoureux2003modeling}, in which a polarizable particle is described in terms of a Core-Drude pair acquiring an induced dipole moment in the presence of an applied electric field and whose evolution is governed by a Langevin dynamics. The model is found to capture key features of the assembly process, such as the early dynamics characterized by the generation of chains of colloids and their clustering into hierarchical structures by forming crossing points among the chains.

The paper is organized as follows. In Section II, we describe the computational model of the polarized particle, and in Section III, we report our numerical results. In particular, we show the early and late time evolution of the assembly process and a cluster analysis which quantitatively captures their formation.

\section{Computational model}\label{R0}

To shed light on the basic mechanisms driving the formation of assembled and ordered structures of nanoparticles, we have run numerical simulations neglecting hydrodynamic interactions and considering only steric interactions, Brownian motion and explicit many-body effects induced by the presence of an external electric field.

\begin{figure}[!h]
\centering
\includegraphics[scale=.5]{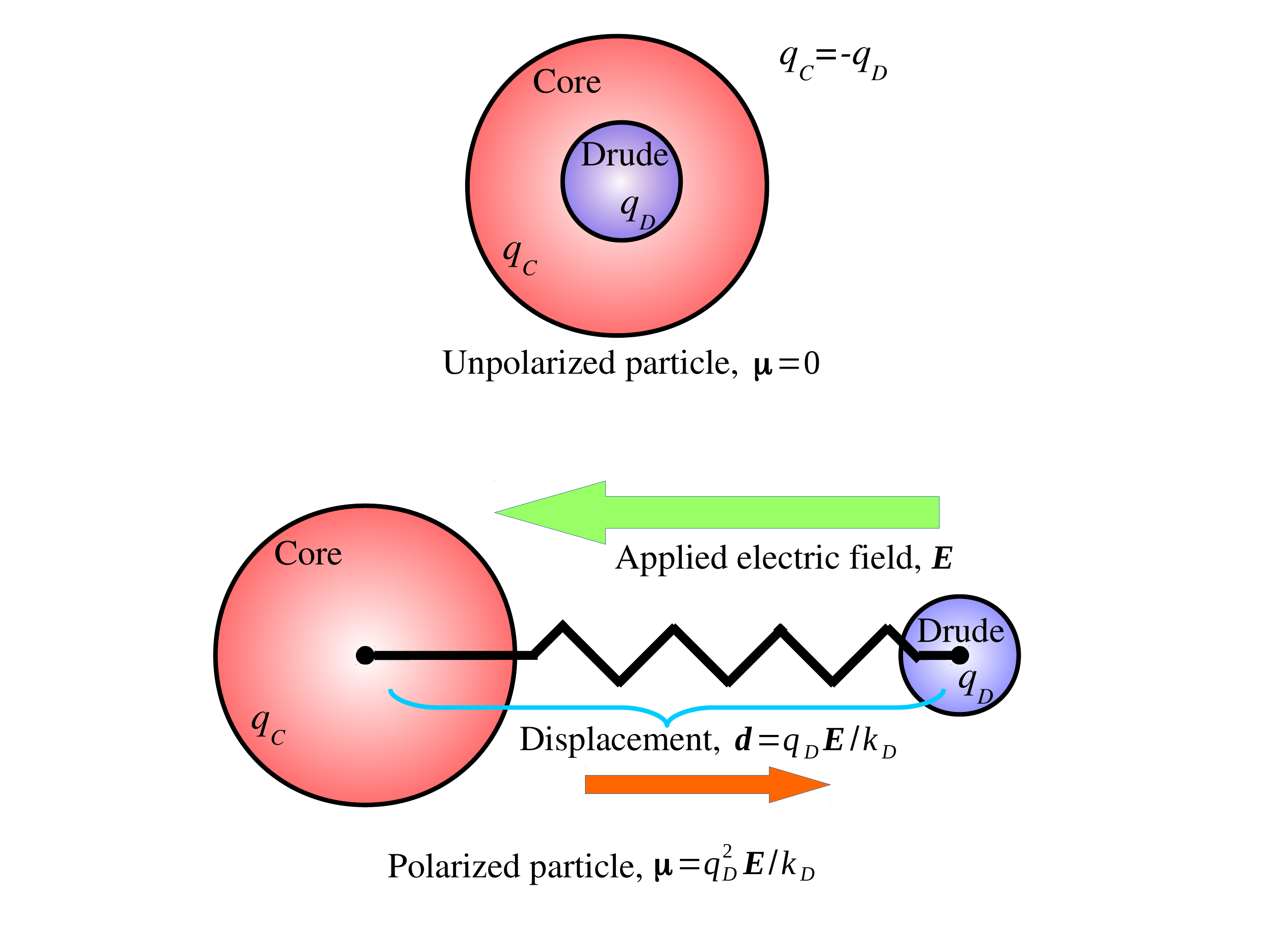}
\caption{The cartoon shows the Core-Drude particle model. In the absence of external electric field ${\bf E}$, the dipole moment is ${\bm \mu}=0$. Here the Core charge is positive the Drude one is negative, with $q_C=-q_D$. Once an electric field is applied, a spring-like force opposes to the field and the dipole moment ${\bm \mu}=q^2{\bf E}^2/k_D$ becomes permanent.}
\label{fig2}
\end{figure}

More precisely, the polarizability $\alpha$ of a particle with a charge $q$ is modeled by introducing a mobile Drude particle (DP) having 
charge $q_D$ bound by a harmonic potential with elastic constant $k_D$ (assumed the same for all bonds)  to a core particle (CP) with 
charge $q_C=q-q_D$ (see Fig.\ref{fig2}). This sets the net charge of the system Drude-Core particle to $q$ \cite{lamoureux2003modeling}, 
in the following assumed equal to zero (neutral charge). In the absence of an electric field, the Drude particle oscillates around the
 equilibrium position ${\bf r}_{C,0}=0$ of the Core particle in the center of mass frame of the  Drude-Core two-particle system. 
Hence, the couple Drude-Core has a net charge $q$ with average dipole moment ${\bm \mu}=0$. If a uniform electric field ${\bf E}$ is applied,
 the Drude is displaced at distance ${\bf d}=q_D{\bf E}/k_D$ in the center of mass frame, and the average 
dipole moment is ${\bm \mu}=q^2_D{\bf E}/k_D$. 
Thus one gets $\alpha=q^2_D/k_D$ \cite{dequidt2016thermalized,lamoureux2003modeling}. 

In the external reference frame and neglecting ${\bf r}_{C,0}=0$, the time evolution of each Drude and Core particle is governed  by two Langevin equations, respectively:
\begin{eqnarray}
  m_C\frac{d^2{\bf r}_{C,i}}{dt^2}&&=-\gamma_C\frac{d{\bf r}_{C,i}}{dt}-\frac{\partial U}{\partial {\bf r}_{C,i}}+\sqrt{2k_BT\gamma_C}\xi_i(t)+q_c{\bf E},\\
    m_D\frac{d^2{\bf r}_{D,i}}{dt^2}&&=-\gamma_D\frac{d{\bf r}_{D,i}}{dt}-\frac{\partial U}{\partial {\bf r}_{D,i}}+\sqrt{2k_BT\gamma_D}\xi_i(t)+q_D{\bf E},
\end{eqnarray}
where ${\bf r}_{C,i}$ and ${\bf r}_{D,i}$ are the positions of the $i$-th Core and Drude particles, $\gamma_C$ and $\gamma_D$ are the viscous friction felt by each of them,  $k_B$ is the Boltzmann constant, $T$ is the temperature, $\xi_i$ is an uncorrelated Gaussian noise with zero mean and unit variance, $m_C$ and $m_D$ (with $m_D\ll m_C$) are the masses of the Drude and the Core particles and ${\bf E}$ is an external electric field. Finally, $U$ represents the total potential given by the sum of three terms $U=U_{bond}+U_{elec}+U_{WCA}$. The first one is the Core-Drude harmonic potential
\begin{equation}
  U_{bond}=\sum_i\frac{1}{2}k_D({\bf r}_{C,i}-{\bf r}_{D,i})^2,
\end{equation}
the second one represents the Coulomb interactions
\begin{equation}
  U_{elec}=\sum_{i}\sum_{j>i}\frac{q_iq_j}{|{\bf r}_i-{\bf r}_j|}
\end{equation}
and the third one is a WCA potential accounting for the interparticle repulsion given by
\begin{equation}
  U_{WCA}=
  \begin{cases}
    4\epsilon [(\frac{\sigma}{r})^{12}-(\frac{\sigma}{r})^6]+\epsilon \hspace{0.3cm} if \hspace{0.3cm} r<2^{1/6}\sigma\\
    0 \hspace{3.15cm} if \hspace{0.3cm} r\ge 2^{1/6}\sigma.
  \end{cases}
\end{equation}
Here $\epsilon$ defines the energy scale, $r$ is the center-to-center separation between Drude and core particles, and $\sigma$ is the value of $r$ at which
$U_{WCA}=0$ which is, to a good approximation, the diameter of the Core particle. It is worth to highlight that the internal Coulomb interaction between the bonded Core and Drude particles is not accounted in the model \cite{dequidt2016thermalized,lamoureux2003modeling}.

\section{Numerical results}
We  have considered a dilute suspension of $322$  colloidal particles of diameter equal to $10$ $\text{nm}$, dispersed in a periodic cubic box of size $3000\times 3000\times 3000$ nm$^3$. This sets a particle volume fraction of $\sim 4\times 10^{-4}$. To capture the dynamic behavior at late times, our simulations are run for $6\times 10^9$ timesteps $\Delta t$, where $\Delta t=2\times 10^{-12}$ s. This corresponds to a total real-time of $12$ milliseconds, which is necessary to catch an almost complete dynamics of the aggregation process. 
%As a consequence, it is worth to note that each simulation is highly demanding in terms of wall-clock time, requiring about three weeks. 
Particle positions are randomly initialized by a uniform distribution with velocities extracted from the Maxwell distribution at a temperature equal to 300 K. We have also set $\alpha=3.8\times 10^{-32}$$m^2/V$~\cite{cao2019measuring}, $m_D/m_C\simeq 1/30$, $\gamma_C=800\times 10^{-9}$s and $\gamma_D=4\times 10^{-9}$s. It is also worth to highlight that the high polarizability $\alpha$ of each colloid is due to the interaction among the colloid itself and the surrounding solvation shell, comprised within a distance of about 10 nanometre \cite{cao2019measuring}. Thus, each colloid is modelled including its respective solvation shell, thus yielding to a spherical object of radius 20 nanometre (the colloidal radius plus the depth of the solvation shell).

We initially discuss the dynamics of the assembly process occurring in the presence of an oscillating electric field ${\bf E}=E_ocos(\omega t){\bf u}_z$, of amplitude $E_o$=0.020 kV/mm and frequency $\nu=\omega/2\pi=500$ kHz, applied along the $z$-axis. Such values are comparable with the experimental ones reported in Ref.\cite{agarwal2009low,kumar2005new}.

%of linear size equal to $5$ $\mu m$.
\begin{figure}[!h]
    \centering
        \includegraphics[width=\textwidth]{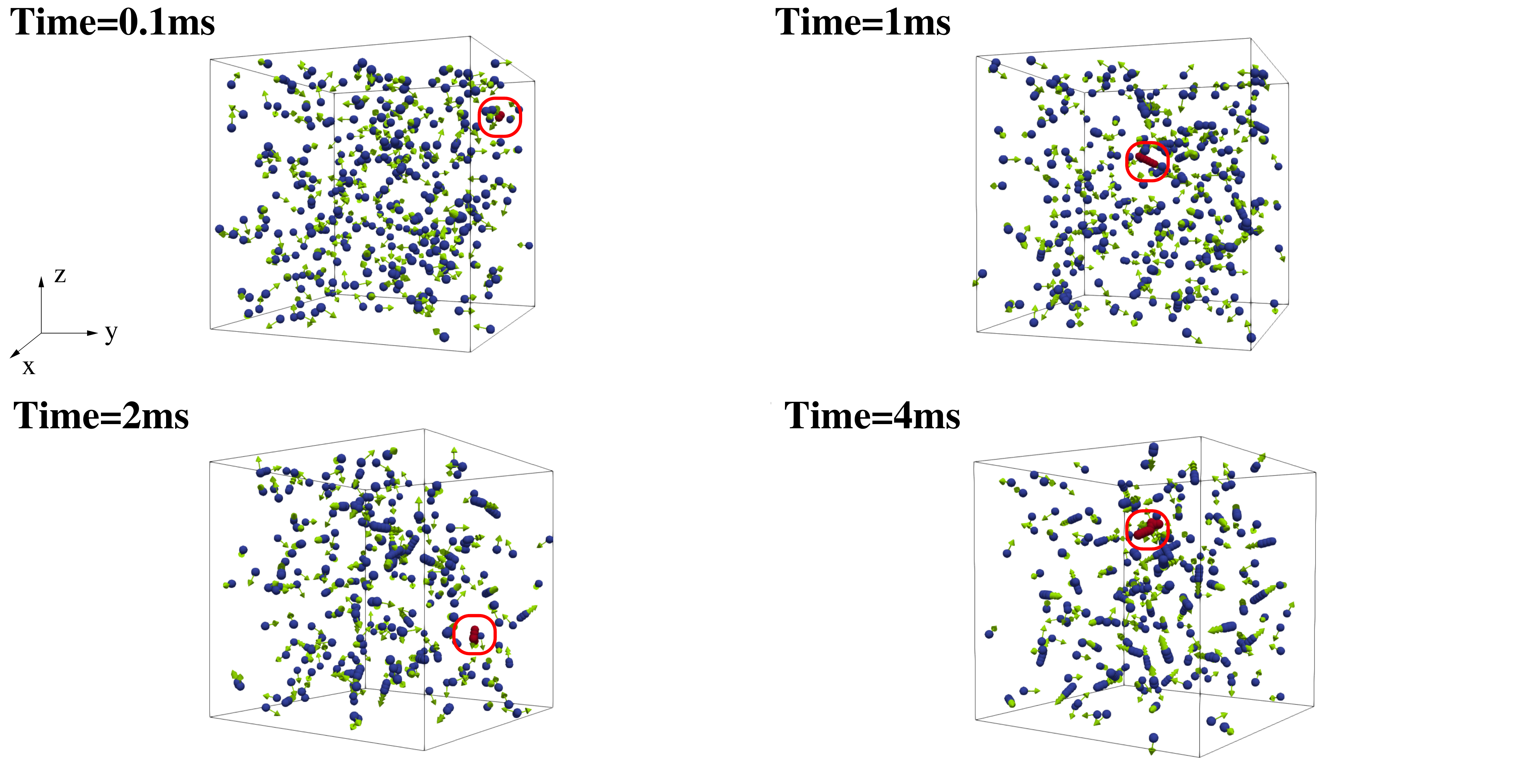}
        \caption{The figure shows the time evolution, up to $4$ms, of the assembly process of nano-particles in the presence of an oscillating electric field. Once the field is switched on, the particles, initially randomly distributed ($t=0.1$ms), gradually approach and assemble into chains ($t=1$ms and $t=2$ms), whose number increases at late times ($t=4$ms). Green vectors indicate the direction of the induced dipole moment. The largest aggregate of colloids is shown in red.}          
    \label{chains}
\end{figure}
%A static electric field of magnitude $|{\bf E}_1|$=0.022 kV/mm is applied along along the $z$-axis along with a second oscillating electric field of magnitude  $|{\bf E}_2|$=0.020 kV/mm and frequency $\nu=500$ KHz.
%The two electric fields increase the particle mobility and accelerate the assembly process.

In Fig.~\ref{chains} a typical early-time evolution of the assembly process is shown. Once the field is switched on, the Core-Drude particles acquire a permanent dipole moment whose orientation fluctuates due to the combined effect of Browian motion and oscillating electric field (Fig.\ref{chains}, $t=0.1$ms). Such dynamics favours the approach of pair of particles which join together due to the attractive Coulomb potential among negatively and positively charged regions (Fig.\ref{chains}, $t=1$ms). This stage is followed by a further one in which such couples, when come close to another particle, assemble and give rise to monodimensional colloidal chains, whose length increases over time (Fig.\ref{chains}, $t=2$ms and $t=4$ms).

\begin{figure}[!h]
\centering
\includegraphics[width=\textwidth]{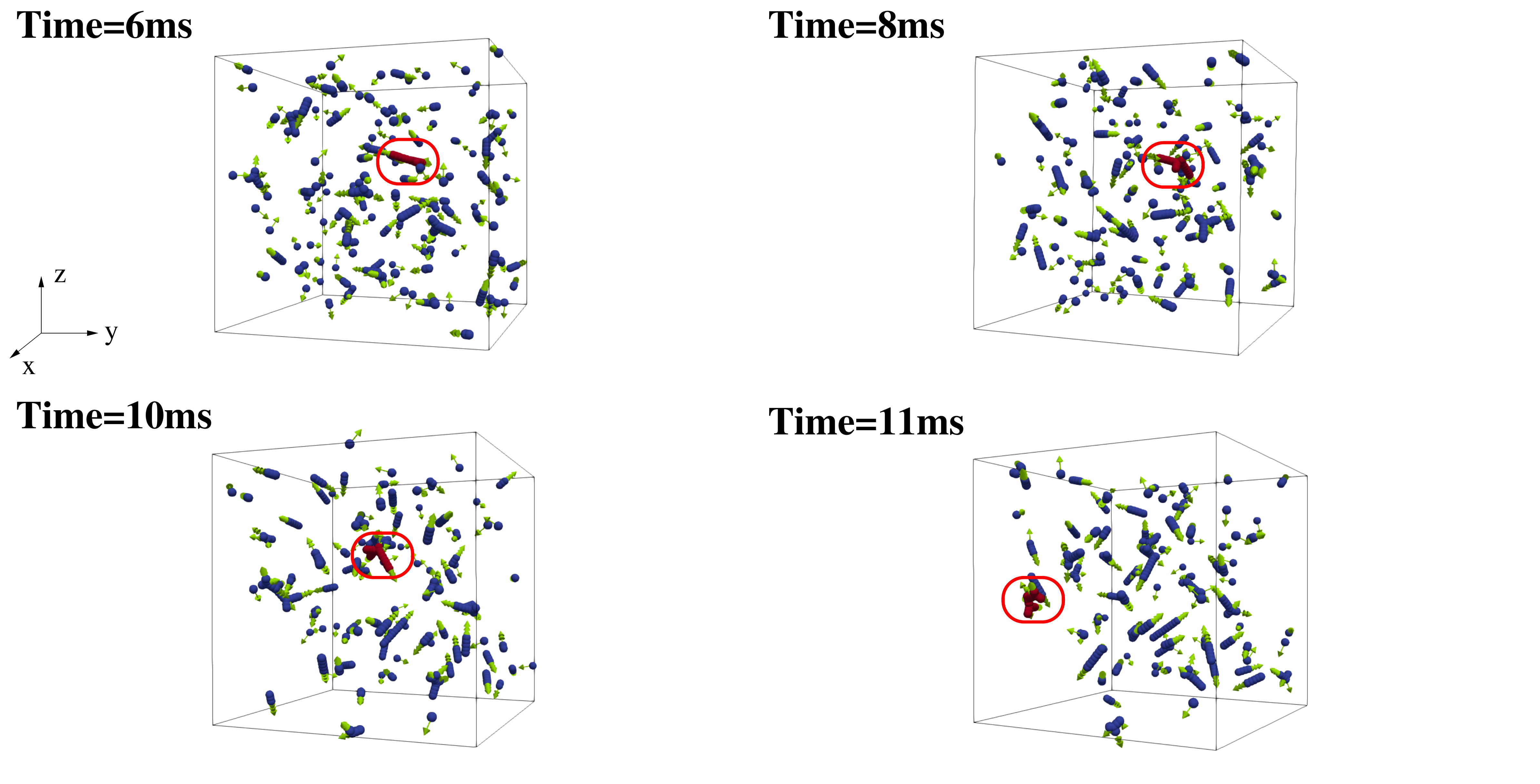}
\caption{The figure shows the incipient formation, occurring from $6$ms to $11$ms, of crossing points among clusters of polarized nanoparticles in the presence of an oscillating electric field. Note that clusters are in average oriented randomly in the simulation box. The vectors indicate the direction of the induced dipole moment. The largest aggregate is shown in red.}\label{cluster_oef}
\end{figure}

The generation of anisotropic chains of colloids resulting from the assembly of polarizable nanoparticles represents the first necessary step preluding to a more complex topolgical arrangement, i.e. the formation of crossing points among chains. This process (whose time evolution is shown in Fig.\ref{cluster_oef}) occurs when colloidal aggregates with different orientations of ${\bm \mu}$ approach and join together, giving rise to fully three-dimensional larger clusters (see Fig.\ref{cluster_oef}, at $t=6$ms and $t=8$ms). Concurrently, the assembly of particle into long chains proceeds and more complex structures, such as three and four-fold aggregates, are produced (see Fig.~\ref{cluster_oef} at $t=10$ms and $t=11$ms), thus increasing the cluster size while diminishing their number.

\begin{figure}[!h]
    \centering
        \includegraphics[width=\textwidth]{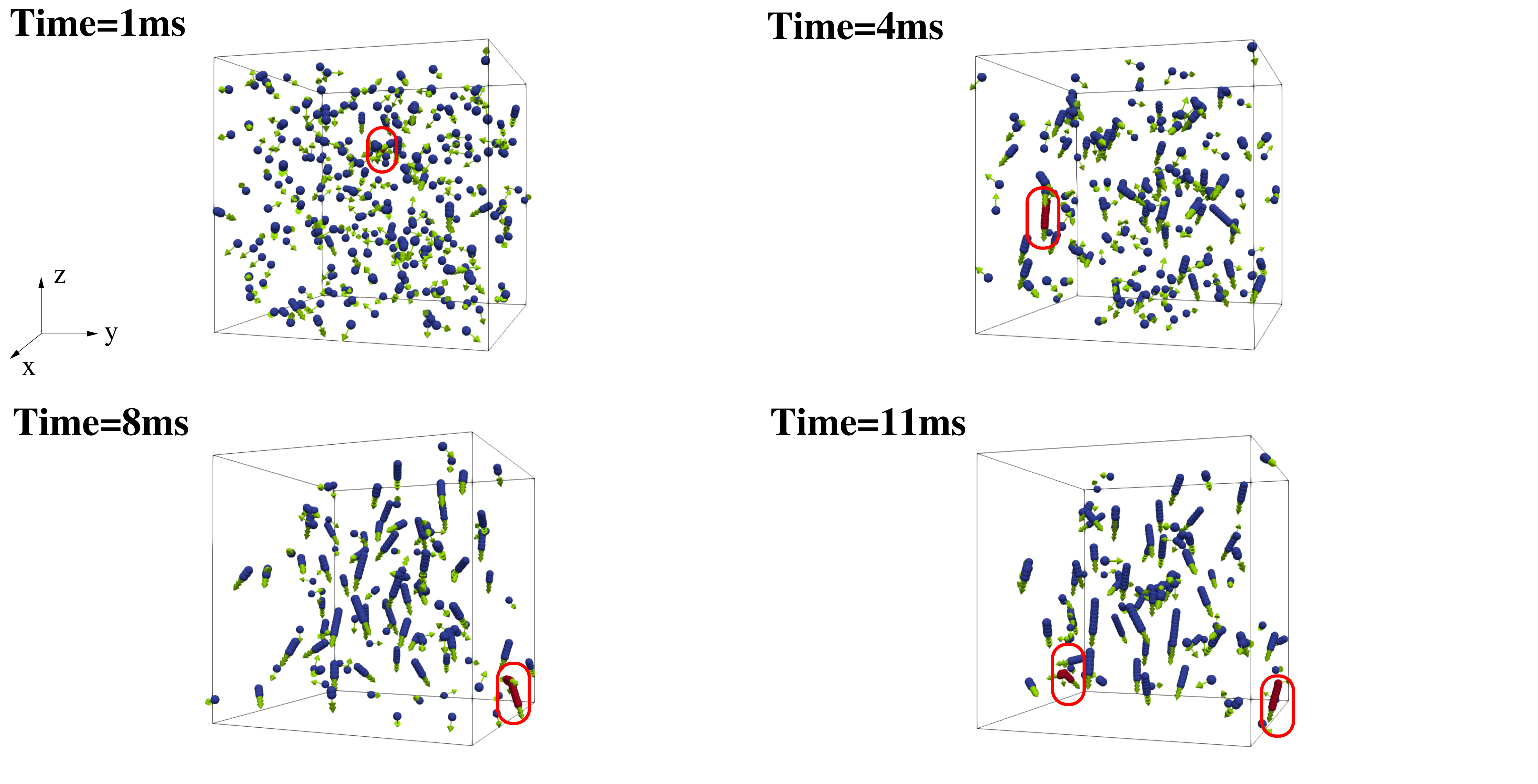}
        \caption{The figure shows the time evolution, up to $11$ms, of the assembly process of nano-particles in the presence of both an oscillating and a static electric field oriented along the $z$-axis. Unlike the case shown in Fig.\ref{cluster_oef}  here, at late times, clusters have a common average alignment parallel to that axis. Green vectors indicate the direction of the induced dipole moment. The largest cluster is shown in red.}
    \label{cluster_sef_oef}
\end{figure}

The addition of a static electric field ${\bf E}_{os}=E_{os}{\bf u}_z$, of magnitude $E_{os}=0.02 kV/mm$ and applied along the $z$-direction, moderately affects the dynamic behavior (see Fig.\ref{cluster_sef_oef}). Indeed, while at early times colloidal aggregates look isotropically distributed within the box ($t=1$ms), subsequently (from t$=4$ms to $t=8$ms) long chains progressively align along the direction imposed by the field and at late times ($t=8$ms) they essentially exhibit a common orientation ($t=11$ms). 

\begin{figure}[!h]
\centering
\includegraphics[width=\textwidth]{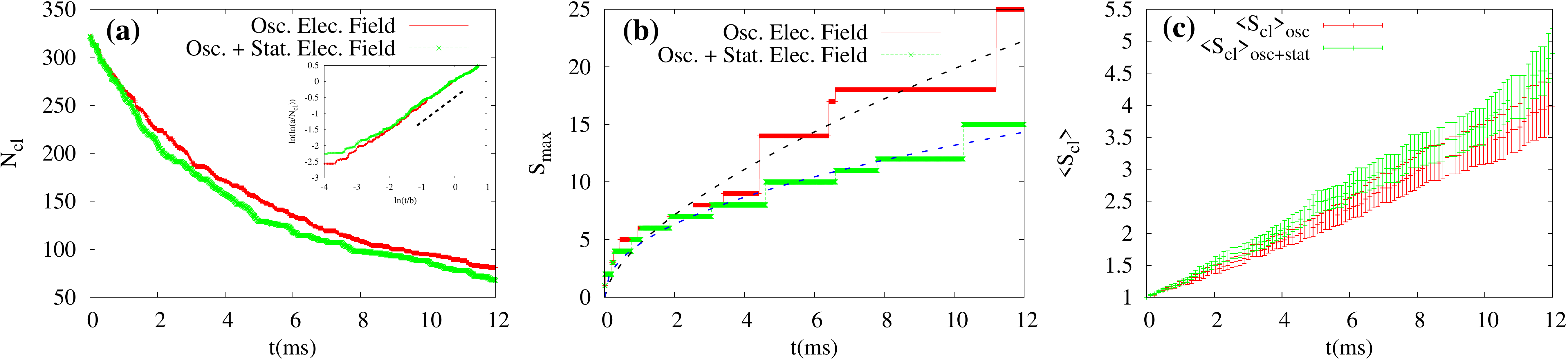}
\caption{(a) Time evolution of number of clusters $N_{cl}$ in the presence of an oscillating electric field (red/plusses) and with an additional static field (green/crosses). In both cases the plots can be best fitted with a stretched exponential $f(t)=ae^{-(t/b)^c}$ where $a\simeq 337$, $b\simeq 7$ $c\simeq 0.72$ for the red/plusse plot , and $a\simeq 348$, $b\simeq 6$, $c\simeq 0.66$ for green/crosses one. The inset shows the linearized curves, while the dotted line represents a guide for the eye with slope $\sim 0.7$. (b) Time evolution of size $S_{cl}$ of the largest cluster of particles. Dashed lines fit the curves with a power law $g(t)=ht^k$, with $h\simeq 4.6$, $k\simeq 0.63$ dashed black line, and $h\simeq 4.6$, $k\simeq 0.45$ dashed blue line. (c) Time evolution of the average cluster size $\langle S_{cl}\rangle$. In the presence of the static field $\langle S_{cl}\rangle$ grows slightly faster.}
\label{fig6}
\end{figure}

A quantitative evaluation of the cluster dynamics is obtained by the Deep First Search (DFS) method \cite{skiena1998algorithm}, which can identify all the colloidal particles belonging to the same cluster, also if they are not directly connected, by using methods of the Graph theory.
Assuming two colloids to be in direct connection whenever the mutual distance is below 50 nm, the cluster analysis of the dynamics shown in Fig.\ref{cluster_oef} and in Fig.\ref{cluster_sef_oef} is reported in Fig.\ref{fig6}. In both cases the number of clusters $N_{cl}$, starting from $N_{cl}=322$ single colloidal particles, diminishes over time (Fig.\ref{fig6}a) due to colloidal aggregation, although the presence of a sufficiently intense static electric field favours a faster decrease. This occurs basically because clusters are forced to orient preferentially along one axis (the $z$-axis), thus having less chance to turn around and catch other neighbouring aggregates. The curves are found to follow a stretched exponential behavior $f(t)=ae^{-(t/b)^c}$ where $c$ is around $0.7$, a behavior probably due to memory effects induced by the electric field in systems made of charged colloids, very likely absent with uncharged particles. 
%smaller than the value of a pure exponential. This may be likely due to memory effects induced by an electric field in systems made of charged colloids whose behavior, unlike uncharged particles, is decisively affected by the field itself.

Alongside the decrease of $N_{cl}$, the size of the clusters augments over time. This is shown in Fig.\ref{fig6}b where the time evolution of the size $S_{max}$ of the largest cluster is plotted. Here $S_{max}=\max_{n_p\in C_i}\{n_p\}$, where $n_p$ is the number of particles assembled in the $i$-th cluster $C_i$. This quantity, starting from $S_{max}=1$ (i.e. a cluster made of a single particle),  is found to increase less rapidly than the one obtained by including a static electric field, a result overall expected since, as previously mentioned, the capture of nearby colloids is a less likely event. Here the time behavior is well fitted by a power law $t^k$, with $k\simeq 0.45$ and $\simeq 0.64$ obtained with and without the static electric field, respectively. The average size of the clusters increases over time as well (Fig.\ref{fig6}c) although, in the presence of a the static electric field, such growth is slightly favoured. This occurs likely because, in that case, the size of largest cluster is smaller than the one obtained with the sole oscillating field. Hence, smaller clusters can capture more particles, thus increasing the value of $\langle S_{cl}\rangle$.

\begin{figure}[!h]
\centering
\includegraphics[width=\textwidth]{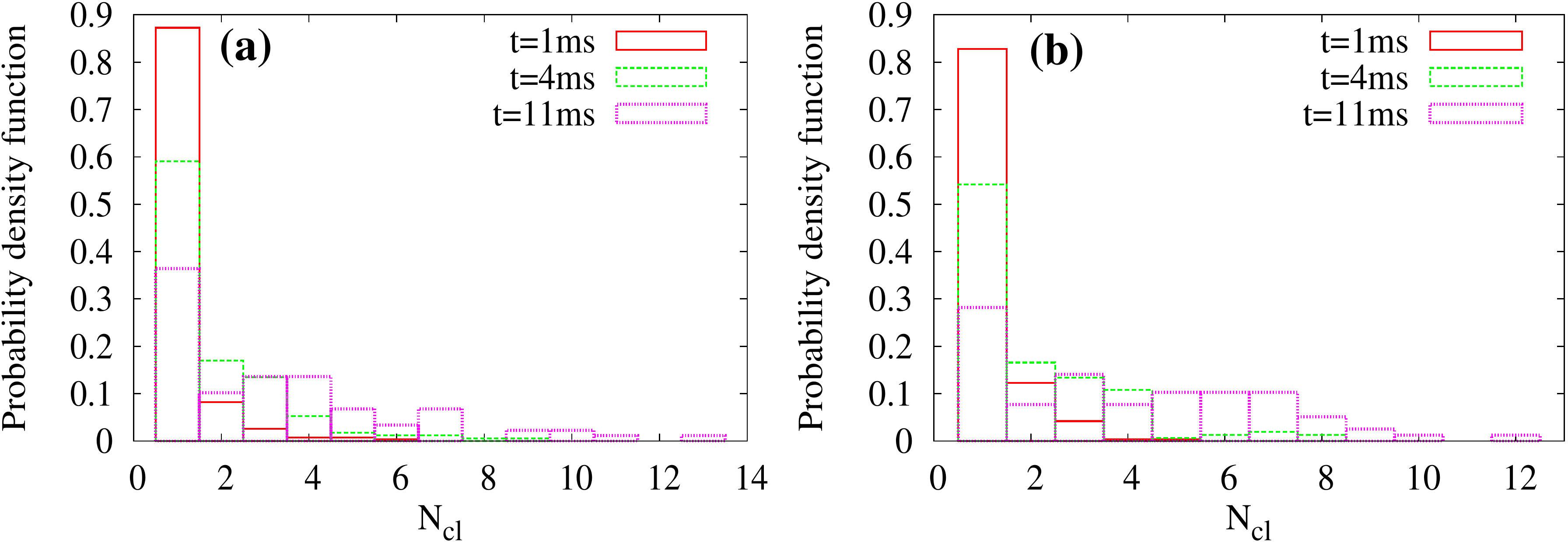}
\caption{Normalized probability density function of the number of colloids $n_c$ at increasing times in the presence an oscillating field (a) and an oscillating plus a static field (b).}
\label{fig7}
\end{figure}

Finally, the cluster dynamics can be also quantified by looking at the time evolution of the normalised probability density function (pdf) of $N_{cl}$ without (c) and in the presence (d) of a static electric field (Fig.\ref{fig7}a-b). Indeed they display a rather similar behavior characterized by a transition from an early-time highly localised distribution towards a broad-shaped one at late times, when $N_{cl}$ exhibits a slightly more uniform structure.

\section{Conclusions and outlook}

To summarize, we have presented a computational scheme for simulating the process of DEP assembly of polarizable particles driven by an external electric field. Particles are modeled by means of a classical version of the Core-Drude theory, in which their dynamics is governed by a Langevin equation.

The model is found to capture key aspects of the assembly process observed experimentally, such as the formation of chain-like aggregates of colloids and the crossing points among different chains. These clusters are the building blocks necessary to achieve more complex arrangements, such as the hierarchical cellular scaffolds described in Ref.~\cite{agarwal2009low,kumar2005new}.

In particular, our simulations show that an oscillating electric field drives the growth of large colloidal clusters, whose number decreases over time following a stretched exponential behavior while the size augments essentially following a power law behavior. The addition of a static electric field with a magnitude akin to the oscillating one, other than favouring a global orientational order of the clusters along the field, produces only a mild effect on the dynamics, slighlty hindering the process of colloidal aggregation. Our results, besides shedding light on the physical mechanisms sustaining the cluster formation, represent a first step towards the modeling of higher complex hierachical arrangements, potentially observed at larger particle volume fraction. 

Further developments of this model will aim at simulating DEP assembly in the presence of hydrodynamic interactions, in order to capture the multiscale physics ranging from particle size level ($\sim$ tens of nanometers) to the typical colloidal domain dimension ($\sim$ tens of micrometers)~\cite{bernaschi2019,montessori2019mesoscale}.\vskip6pt

\section*{Acknowledgments}
The authors acknowledge funding from the European Research Council under the European Union's Horizon 2020 Framework Programme (No. FP/2014-2020) ERC Grant Agreement No.739964 (COPMAT).
The computational project PRACE 16DECI0017 RADOBI is gratefully acknowledged.

\end{document}